\documentclass[10pt,aps,prc,floatfix,nofootinbib]{revtex4-1}   

\usepackage{color}
\usepackage[pdfencoding=auto, pdfpagelabels]{hyperref}
\definecolor{linkcolor}{rgb}{0,0,0.40} 
\hypersetup{%
    pdfsubject=Paper,
    pdfkeywords={nuclear physics} {Bayesian} {EFT},
    unicode = true,
    breaklinks = true,
    colorlinks = true,
    linkcolor = linkcolor,
    citecolor = linkcolor,
    menucolor = linkcolor,
    urlcolor = linkcolor
}

\usepackage{amsmath}
\usepackage{graphicx}
\usepackage{caption}
\usepackage{array}
\usepackage{booktabs}
\newcolumntype{L}{>{$}l<{$}}

\DeclareMathOperator{\pr}{pr}
\newcommand{\given}{\,|\,}  
\newcommand{\yexp}{y_{\text{expt}}}
\newcommand{\yth}{y_{\text{th}}}
\newcommand{\ytrue}{y_{\text{true}}}
\newcommand{\sigmaHO}{\sigma_{\text{H.O.}}}
\newcommand{\sigmaparam}{\sigma_{\text{param}}}
\newcommand{\avec}{\mathbf{a}}
\newcommand{\abar}{\bar{a}}
\renewcommand{\pr}{\text{pr}}

\begin{document}
\title{Neglecting correlations leads to misestimated model errors in EFT predictions}
\author{N.~L. Carter\footnote{nathan.carter@gwu.edu}}
\affiliation{Institute of Nuclear \& Particle Physics and Department of Physics \& Astronomy, Ohio University, Athens, OH 45701, USA
}
\affiliation{Institute for Nuclear Studies, George Washington University, Washington, DC , USA}
\author{R.~J. Furnstahl\footnote{furnstahl.1@osu.edu}}
\affiliation{Department of Physics, The Ohio State University, Columbus, OH 43210, USA}
\author{J.~A. Melendez\footnote{melendez.27@osu.edu}}
\affiliation{Department of Physics, The Ohio State University, Columbus, OH 43210, USA}
\author{D.~R. Phillips\footnote{phillid1@ohio.edu}}
\affiliation{Institute of Nuclear \& Particle Physics and Department of Physics \& Astronomy, Ohio University, Athens, OH 45701, USA
}

\begin{abstract}
 Bayesian analyses of the convergence pattern of Effective Field Theories (EFTs) enable estimation of the uncertainty induced by a truncated expansion. When an EFT that has been calibrated to data is used to make a prediction this truncation uncertainty enters the posterior predictive distribution twice: directly from the finite-order calculation of the predicted quantity and indirectly through the posterior probability distributions of the EFT low-energy constants (LECs) determined by the calibration. In this work, we focus on the interplay of these two sources of uncertainty. We do this in the context of a toy EFT that we fit to pseudodata and use to make predictions. Direct EFT truncation uncertainty and LEC uncertainty are correlated in predictions when the predicted quantity is correlated with the observables used to fit the LECs. Here this results in the overall theoretical uncertainty in the EFT prediction being smaller than either the uncertainty induced by the truncation error or that stemming from the LECs alone.
\end{abstract}


\maketitle

\section{Introduction and Motivation}

Series expansions in a (presumed) small parameter, such as those encountered in perturbative quantum field theories, have an inherent uncertainty when they are calculated to a finite order. 
An additional layer of complexity arises in effective field theories (EFTs), where the coefficients of terms in the series---the low-energy constants (LECs) of the EFT---have to be estimated from data. Truncation uncertainty due to the choice of a particular order of the EFT result for the data being described should be taken into account in estimates of these parameters~\cite{Schindler:2008fh,Brynjarsdottir:2014}.
This is an example of ``model uncertainty'', which in the context of a well-formulated EFT 
is the ``truncation error'' of the EFT (or, more generally, QFT) calculation. 
In this context, a Bayesian approach to LEC parameter estimation
can account for both theoretical expectations regarding the accuracy of the perturbative series and a statistical characterization of  experimental data. Such an approach therefore provides more realistic inference of the parameters of an imperfect theory---and in particular their uncertainties---from the available data. 

The conventional method for estimating the LECs is to minimize a $\chi^2$ objective function for a given set of observed data, $D$. 
Typically, the uncertainties in the $\chi^2$ function are taken to be the experimental errors associated with each observation, although 
the EFT truncation error 
can also be incorporated in the $\chi^2$ function~\cite{Schindler:2008fh,Wesolowski:2018lzj}. 
In a number of publications, including Refs.~\cite{Carlsson:2015vda,Somasundaram:2023sup,Hu:2021trw}, the total uncertainties that appear in the objective function are the sum-in-quadrature of the experimental uncertainty, the uncertainty from the numerical method employed to evaluate the model, and the model uncertainty. Such an objective function assumes that all sources of uncertainty are uncorrelated/independent.

In the prediction of some future data, $z$, given existing data, $D$---i.e.\ the posterior predictive distribution (ppd) of $z$---the uncertainties on the parameters ${\bf{a}}$ must be propagated to the final prediction $z$. 
The model, $M$, being used to predict $z$ is generally the same imperfect model (for example, a particular EFT, truncated at a specific order) used to estimate the parameters ${\bf a}$. 
There is thus a direct impact of model uncertainty on the probability density function (pdf) of $z$ because the impact of its truncated expansion should  be accounted for in the ppd. Meanwhile, the parametric uncertainties represented in the pdf of ${\bf{a}}$, if assessed in the presence of model uncertainty, will constitute a second, indirect, impact of model uncertainty on the pdf of $z$. 

Few nuclear physics predictions include both the full parametric uncertainty and the direct model uncertainty; those which do combine them assuming their independence, i.e., in quadrature---see, e.g., Ref.~\cite{Hu:2021trw}. However, combining the model and parametric uncertainty under this assumption may substantially misestimate the total uncertainty. It runs the risk of double counting the model uncertainty that was accounted for when the posterior of the parameters $\avec$ was obtained. We note that neglecting correlations between theoretical uncertainties has been shown to lead to overestimated uncertainties in other contexts, e.g., the nuclear symmetry energy~\cite{Drischler:2020yad} and the energy spectra of p-shell nuclei~\cite{Maris:2020qne}. 

In this paper, we manifest the interplay of parametric uncertainty and truncation/model uncertainty using the toy model of EFT expansions first proposed in Ref.~\cite{Schindler:2008fh} and explored further in Refs.~\cite{Wesolowski:2015fqa,Connell:2021qcd}. The model, together with the Bayesian formalism needed for our work, is reviewed in Sec.~\ref{sec:theory}. 
We introduce the ``naturalness priors'' on the coefficients in the perturbative series and review the use of marginalization to calculate posterior probability distributions for both the coefficients in the series and predictions of the value of the underlying function $g(x)$ at some new point in the input space $x$. 
In Sec.~\ref{sec:predictions} we implement the formulae derived in Sec.~\ref{sec:theory} and show that the final uncertainty on predictions is markedly smaller than either the parametric uncertainty or the (unconditional) truncation uncertainty. 
We summarize and suggest some implications for the application of EFTs in nuclear physics in Sec.~\ref{sec:conclusions}.

\section{Theory}
\label{sec:theory}

\subsection{Statistical model and posterior predictive distribution}

Let us denote the underlying true theory (what statisticians call ``truth'') as $\ytrue(x)$, where $x$ is a generic input (i.e., it could be a vector in the input space).
Then $\ytrue$ is related to both the observed data and the theoretical model.
Suppose we have calibrated the parameters $\avec$ using a set of data $\{y_i\}$ taken at inputs $\{x_i\}$. 
Then $y_i$ at a particular input $x_i$ should be the truth plus the experimental error $\delta \yexp$:
\begin{equation}
     y_i = \ytrue(x_i) + \delta\yexp(x_i) .
    \label{eq:expt_truth}
\end{equation}
At the same time, the truth should be our model predictions plus the theory error (model discrepancy):
\begin{equation}
    \ytrue(x_i) = \yth(x_i;\avec) + \delta\yth(x_i) ,
    \label{eq:theory_truth}
\end{equation}
where we have explicitly noted that the model predictions depend on parameters $\avec$ but the theory error does not. 
Eliminating $\ytrue$ yields our statistical model for the observations~\cite{kennedy2001bayesian,Brynjarsdottir:2014},
\begin{equation}
    y_i = \yth(x_i;\avec)  + \delta\yexp(x_i) + \delta\yth(x_i) .
    \label{eq:stat_model}
\end{equation}
Rather than an equation relating purely deterministic functions, Eq.~\eqref{eq:stat_model}
encodes the relationship between probability distributions for the random variables $y_i$, $\yth(x_i;\avec)$, $\delta\yexp(x_i)$ and  $\delta\yth(x_i)$. This defines a posterior probability distribution function (pdf) for observed data.

We then consider the 
prediction of a datum $z=\ytrue(x)$ at a point $x$ that is not part of the original set $\{x_i\}$.
If we do not fold anticipated experimental error into our prediction, then the relationship between $z$ and the theory model result $\yth(x;\avec)$ is given by \begin{equation}
    z=\yth(x;\avec) + \delta \yth(x).
\end{equation}
The Bayesian expression for the prediction of $z$ given the observed data $D=\{y_i\}$ is then the posterior predictive distribution (ppd),
\begin{equation}
    \pr(z \given D, I) = \int \delta\bigl(z - \yth- \delta \yth(x)\bigr)\,
    \pr(\yth,\delta \yth(x) \given D,I)\, d\yth d \delta \yth(x).
    \label{eq:ppdfull}
\end{equation}
Here $I$ stands for additional information we are not making explicit (such as the theoretical model $M$), while $\pr(A \given B, I)$ is read as `the probability that $A$ is true given that $B$  and $I$ are true'. Note that at this stage we have a placeholder random variable $\yth$ standing in for $\yth(x;\avec)$, since we have not made all dependencies explicit in the integral yet. 

We want to consider two separate limits of Eq.~(\ref{eq:ppdfull}). First, suppose that the theory discrepancy $\delta \yth(x)$ is independent of $\yth$, and is not constrained by the data. In that case, the pdf in the integrand of Eq.~(\ref{eq:ppdfull}) factorizes, and we have
\begin{equation}
    \pr(z \given D, I) = \int \delta\bigl(z - \yth - \delta \yth(x)\bigr) \,
    \pr(\yth \given D,I) \, \pr(\delta \yth(x) \given I) \, d\yth d\delta\yth(x).
    \label{eq:ppdp}
\end{equation}
Here we identify the ppd for $\yth$, $\pr(\yth \given D,I)$. Crucially, this is a different ppd than the full ppd we are computing, that for $z=y(x)$. The ppd for $\yth$ is given by the usual marginalization over the parameter posterior $\pr(\avec\given D,I)$:
\begin{equation}
    \pr(\yth \given D,I)=\int  \pr(\yth(x;\avec) \given \avec,I)\, \pr({\bf a} \given D,I)\, d\avec,
\end{equation}
where we have used the fact that the model for $\yth(x;\avec)$ is deterministic and so $\pr(\yth(x) \given {\bf a},I)$ is not conditional on the data $D$, because it is a delta function that is determined completely by the functional form of the model $\yth(x;\avec)$.

However, if $\delta \yth(x)$ and $\yth(x;\avec)$ are jointly conditioned on the data, as will happen if the theory discrepancy for $y(x)$ is fit together with the parameters $\avec$, then the pdf, and hence the integral \eqref{eq:ppdfull}, does not factorize. In that case marginalizing in the parameters yields:
\begin{equation}
    \pr(z \given D, I) = \int \delta\bigl(z - \yth(x;\avec) - \delta \yth(x)\bigr)
 \,
\pr(\avec \given \delta \yth(x),D,I) \pr(\delta \yth(x) \given D,I) \, d\avec\, d \delta\yth(x).
\label{eq:ppdwithdiscrepancy}
\end{equation}
In this case, if we wish to correct the ppd for $\yth$ by the theory uncertainty in order to construct the ppd for $z$ then we need to sample $\delta \yth$\footnote{For linear models with Gaussian priors, like the toy model described in this paper, this could also be done analytically following the Bayesian linear model.}. Then, for each value of the model discrepancy term that is considered in the integral over $\delta \yth(x)$, the ppd of $\yth(x)$ can be combined with $\delta \yth(x)$ to form $z$.

In the joint ppd $\pr(\avec,\delta \yth|D,I)$ the quantities $\avec$ and $\delta \yth$ are not independent because Eq.~(\ref{eq:stat_model}) determines this pdf. Provided $\delta y_{\rm expt}(x_i)$ is small compared to $\delta \yth$ this relationship forces the effect of parametric and truncation uncertainty on $\yth$ to be anti-correlated.
Not properly accounting for this anti-correlation can lead to overestimated errors in the prediction for $z$. 

The formulation presented here is quite general, but it is also somewhat abstract. It is easier to see how this plays out in practice in the simple model we now present. 

\subsection{The model problem}
\label{sec:problem}

We employ the toy model of an EFT expansion first discussed in Ref.~\cite{Schindler:2008fh}. That work considered an underlying function 
\begin{equation}
g(x) = \Bigl[\frac{1}{2} + \text{tan}\bigl(\frac{\pi}{2}x\bigr)\Bigr]^{2} ,
\label{eq:underlying}
\end{equation}
which is constructed to have a radius of convergence of one for its Taylor series. 
While an EFT expansion may be asymptotic, in practice one truncates well before the divergence occurs so modeling it by a convergent Taylor series is reasonable.
This function will be used here as the data generating process both for the training data used in parameter estimation and the evaluation data used to study predictions.

The ``Effective Field Theory'' of $g(x)$ for $x < 1$ is a finite-order Taylor expansion of the function. The $k$th-order EFT expression takes the form:
\begin{equation}
{\yth}_{k}(x;\avec) = a_{0} + a_{1}x + a_{2}x^{2} + ... + a_{k}x^{k}.
\label{eq:EFTOk}
\end{equation}
We know the true values of these coefficients are $a_0= 0.25$, $a_1= 1.65$, $a_2=2.99$, $a_3=0.32$, and $a_4= 0.06$ since we chose the ``underlying theory'' $g$. 

However, in practical applications we typically do not have access to $g$ itself. Instead we must learn about the coefficients (LECs) in Eq.~(\ref{eq:EFTOk}) by fitting data on the observable that $g$ corresponds to. To simulate this situation we generate a data set $D$ of $N$ values $\{y_i\}$ by adding a 5$\%$ random offset $\delta \yexp$ (simulating experimental error) to the true value of $g$ at a set of points $\{x_i:i=1,\ldots,N\}$, so:
\begin{eqnarray}
    y_i&=&g(x_i) + \delta \yexp(x_i), 
    \label{eq:toy_data} \\ 
    \delta \yexp(x_i) &\sim& {\cal N}(0,[0.05 g(x_i)]^2).
    \label{eq:datageneration}
\end{eqnarray}
But ${\yth}_k(x;\avec)$ also differs from $g(x)$ by a truncation error, which is of $O(x^{k+1})$. We can therefore write:
\begin{equation}
    g(x)={\yth}_{k}(x;\avec) + \delta {\yth}_{k}(x),
    \label{eq:gfk}
\end{equation}
where $\delta {\yth}_k(x)$ is the model discrepancy for the order $k$ polynomial. 
Note that Eqs.~\eqref{eq:toy_data} and \eqref{eq:gfk} are the manifestations in the toy model of Eqs.~\eqref{eq:expt_truth} and \eqref{eq:theory_truth}, respectively.

In contrast to many situations in which mis-modeling is considered, in EFTs the parametric form of the model discrepancy is specified by the identification of that model discrepancy with a higher-order term (or terms).
This is modeled as a random variable, here distributed as a Gaussian with mean zero and variance due to higher-order terms up to some order $k'$, so that 
\begin{equation}
    \delta {\yth}_k(x) \sim {\cal N}\left(0, \sum_{n=k+1}^{k'} \abar^2 x^{2n}\right).
    \label{eq:deltaythnat}
\end{equation}
Moreover, the EFT requirement of naturalness means that we anticipate
$\abar$ is of order one. Here, $\abar$
is a hyper-parameter that can be assigned its own prior, or can be taken to have a fixed value that is large enough that reasonable values of the pre-factor of the $O(x^{k+1})$ uncertainty are accommodated in the analysis.

\subsection{Applying Bayes' Theorem}
\label{sec:Bayes}

%

The joint posterior for the parameters $\avec$ and the higher-order coefficients $\avec_{\text{trunc}} \equiv (a_{k+1},\ldots,a_{k'})$ is obtained using Bayes' theorem\footnote{We have dropped the marginal likelihood that typically appears in the denominator because we will normalize the posterior pdf of parameters to one.}:
\begin{equation}
    \pr({\bf a}, {\bf a}_{\text{trunc}} \given D, k,k',I) \propto \pr(D \given {\bf a},{\bf a}_{\text{trunc}},k,k',I) \pr({\bf a} \given k, I) \pr({\bf a}_{\text{trunc}} \given k,k',I).
    \label{eq:ppdavec}
\end{equation}

 The form of the first factor, $\pr(D \given \avec,{\bf a}_{\text{trunc}},k,k',I)$, is determined by the data-generation process (\ref{eq:datageneration}) and the discrepancy between ${\yth}_k$ and $g$. 
 We assume that the noise in the different measurements $y_i$ are independent and uncorrelated across the input space. (This assumption can be relaxed if desired.) Then all we need are the  variances of the ``experimental data":
 \begin{equation}
    \sigma_i^2=(0.05 g(x_i))^2.
 \end{equation}
 The likelihood $\pr(D \given {\bf{a}},{\bf a}_{\text{trunc}},k,k',I)$ is then
\begin{equation}
    \pr(D \given {\bf{a}},{\bf{a}}_{\text{trunc}},k,k',I) \propto \exp(-\chi^2/2),
    \label{eq:likelihood}
\end{equation}
with
\begin{equation}
\chi^{2} = \sum_{i = 1}^{N}\Bigl(\frac{y_{i} - {\yth}_k(x_{i};\avec)-\delta {\yth}_k(x_i;{\bf a}_{\text{trunc}})}{\sigma_{i}}\Bigr)^{2},
\label{eq:chisq}
\end{equation}
where 
\begin{equation}
    \delta {\yth}_k(x;{\bf a}_{\text{trunc}})=\sum_{j=k+1}^{k'} a_j x^{j}
    \label{eq:deltaythexplicit}
\end{equation}
is the model discrepancy, computed up to some order $k' > k$~\footnote{The higher-order coefficients $a_{k+1},\ldots,a_{k'}$ can be marginalized over here and this result reformulated in terms of a theory covariance matrix that is combined in quadrature with the diagonal experimental covariance matrix~\cite{Wesolowski:2018lzj}.}.
%
To specify the second term in \eqref{eq:ppdavec} we adopt a Gaussian form for the prior on the parameters and the coefficients $\avec_{\text{trunc}}$:
\begin{equation}
\pr({\bf a},\avec_{\text{trunc}}\given I) = \text{exp}\left(-\frac{\textbf{a}^{2}+\avec_{\text{trunc}}^2}{2\abar^{2}}\right).
\label{eq:parameterprior}
\end{equation}
The value of $\abar$  determines the extent to which the parameters $\avec$ can excurse from their (assumed) mean value of zero. 
By keeping the value of $\abar^{2}$ of $\mathcal{O}(1)$, we penalize values of the parameters that are unnaturally large, ensuring the naturalness condition is incorporated. 

 The likelihood and prior of Eqs.~\eqref{eq:likelihood} and \eqref{eq:parameterprior} yield the log posterior pdf 
\begin{equation}
\ln(\pr(\textbf{a},\avec_{\text{trunc}} \given  D,k,k',I)) = - \frac{1}{2}\sum_{i = 1}^{N}\biggl[\Bigl(\frac{y_{i} - {\yth}_k(x_{i};\textbf{a})-\delta {\yth}_k(x_i;{\bf a}_{\text{trunc}})}{\sigma_{i}}\Bigr)^{2} + \frac{\textbf{a}^{2}+{\bf a}_{\text{trunc}}^2}{\abar^{2}}\biggr] + \ldots,
\label{eq:posteriorwprior}
\end{equation}
for the parameters $\textbf{a}$ of the polynomial model of Eq.~(\ref{eq:EFTOk}).
Note that by taking $\abar^{2}$ to $\infty$, the prior term goes to zero, and we return the conventional likelihood and $\chi^{2}$ of Eqs.~\eqref{eq:likelihood} and \eqref{eq:chisq}.

\subsection{The effect of higher orders and marginalization on parameter estimation}
\label{sec:parameters}

Marginalization integrates over a subspace of parameters to obtain a pdf of only the parameters we are interested in while accounting for the impact of the marginalized parameters~\cite{Sivia:2006,Schindler:2008fh}. 
For example, to get the one-dimensional pdf of the parameter $a_{j}$ we marginalize over the probability spaces for all $a_{i}$, excluding the parameter of interest, but including the coefficients of the terms in the discrepancy $\delta {\yth}_k$:
\begin{equation}
\pr(a_{j} \given  k,I) = \int \pr({\bf a} \given  k,I) \, da_{0} \ldots da_{j-1} da_{j+1} \ldots da_{k} da_{k+1} \ldots da_{k'}.
\label{eq:marginalizeto1d}
\end{equation}

In this work, we need to know how higher-order uncertainty affects the estimates of the parameters $\avec$ that enter the function ${\yth}_k(x;\avec)$. 
Accounting for this uncertainty in the estimates of $a_0$, $a_1$, $a_2$, \ldots, $a_k$ can be accomplished by estimating the parameters for $a_{k+1}, \ldots, a_{k'}$ (with $k' > k$), and marginalizing over them. 
We note that this split between the parameters we are interested in (the $a_j$'s for $j \leq k$) and the higher-order coefficients (the $a_j$'s for $j > k$) is essentially arbitrary, since in this example the discrepancy function associated with the contributions from orders $k+1$ to $k'$ has the same form as the function itself. Therefore in this toy example (although not in the general case encountered in EFTs) discrepancy modeling of the higher-order terms could equally be thought of as just doing parameter estimation at a higher order.

We carry out parameter estimation for $k=3$ and $k'=4$ and $5$ with  the data set $D$ being the set $D_1$ from Refs.~\cite{Schindler:2008fh,Wesolowski:2015fqa}. It is listed in Table~\ref{tab:toy_data} and consists of 10 equally spaced points in $x$ between 0 and $1/\pi$ with the noise assigned to these pseudo-data as 5\% of the output value as in Eq.~\eqref{eq:datageneration}.

\begin{table}[tb]
\renewcommand{\arraystretch}{1.2}
\centering
\caption{Experimental Data}
\label{tab:toy_data}
\begin{ruledtabular}
\begin{tabular}{ccc}
$x \cdot \pi/2$ & $y$ & $\sigma$ \\
\midrule
0.05 & 0.31694 & 0.01585 \\
0.10 & 0.33844 & 0.01692 \\
0.15 & 0.42142 & 0.02107 \\
0.20 & 0.57709 & 0.02885 \\
0.25 & 0.56218 & 0.02811 \\
0.30 & 0.68851 & 0.03443 \\
0.35 & 0.73625 & 0.03681 \\
0.40 & 0.87280 & 0.04364 \\
0.45 & 1.0015 & 0.0501 \\
0.50 & 1.0684 & 0.0534 \\
\end{tabular}
\end{ruledtabular}
\end{table}

We set the hyperparameter $\abar=5$. Results were obtained using the MCMC sampler emcee~\cite{Foreman_Mackey:2013aa}, with 500 burn-in steps and 10,000 steps for 50 walkers. The auto-correlation time was 150 samples. 
We obtain the means and 68\% intervals displayed in Table \ref{table:parameters}. These are consistent with the results of previous studies~\cite{Schindler:2008fh,Wesolowski:2015fqa}: for example, we observe that the uncertainty on $a_0$ is quite stable for all choices of $k$ and $k'$. We note that since this is a linear model with Gaussian priors and a Gaussian likelihood the results can also be obtained analytically. 

\newcommand{\st}{\rule[0.1cm]{0pt}{0.3cm}}
\begin{table*}[htb!]
        \caption{Parameters $a_0, a_1, a_2, a_3$ from fits up to $k = 3$ with: no model discrepancy ($k'=3$, first row), one term only in model discrepancy ($k'=4$, second row), two terms in model discrepnacy ($k'=5$, third row). Note that the error bars should be symmetric, the small asymmetry is due to the use of a finite number of samples.}

    \setlength{\tabcolsep}{8pt}
        \begin{tabular}{L|L|L|L|L|L|}
             k & k' & a_{0} & a_{1} & a_{2} & a_{3}\\
             \hline
              3 & 3 & 0.25^{+0.02}_{-0.02} \st & 1.64^{+0.45}_{-0.44} & 3.0^{+2.3}_{-2.3} & 0.35^{+4.3}_{-4.3}\\
             3 & 4 & 0.25^{+0.02}_{-0.02} \st & 1.65^{+0.45}_{-0.45} & 2.99^{+2.3}_{-2.4} & 0.32^{+4.4}_{-4.3}\\ 
             3 & 5 & 0.25^{+0.02}_{-0.02}\st & 1.65^{+0.45}_{-0.45} & 2.95^{+2.3}_{-2.4} & 0.32^{+4.4}_{-4.3} \\
        \end{tabular}
        \label{table:parameters}
\end{table*}

Not shown in  Table \ref{table:parameters} are the results for the ``nuisance" parameters $a_{k+1}, \ldots, a_{k'}$ that parameterize the model discrepancy. 
The median value of $a_{k'}$ approaches 0 as $k'$ increases: for the $k'=4$ case we have $a_4=0.06 \pm 4.9$ and for $k'=5$, $a_5=0.02 \pm 5$. This is expected if the pdf of this higher-order coefficient is dominated by the prior. When $k'=5$ the size of the 68\% interval for $a_4$ and $a_5$ is determined by $\abar$~\footnote{Both in previous work and in our study we tested other values of $\abar$ and found that the 68\% interval scaled with $\abar$. The analytic result also demonstrates this.}.
At least as far as the one-dimensional posteriors quoted here are concerned, the likelihood does little to constrain $a_4$ and $a_5$, with their posterior essentially ``returning the prior". This is likely because the calibration of ${\bf a}$ is done at values of $x$ that are small compared to the radius of convergence of the Taylor series. If we were closer to the radius of convergence, and the data were sufficiently precise there, we would see stronger signals of non-zero coefficients at higher order. 

In the following section we will consider the fourth-order term as dominating the truncation uncertainty, and the cubic ($k=3$) as the function whose parameters we are estimating. 
But how much does the $k' = 4$ term actually change the parameter estimation? Table \ref{table:parameters} shows that the $a_{3}$ and $a_{2}$ central values change by 10\% and 1\% respectively in the fourth-order compared to the third-order fit. This is well within the (truncation-error-broadened) 68\% intervals for these quantities. 
Once we are fitting to polynomials of these degrees, the fit to these data, which only extend up to $x=1/\pi$, remains quite stable as higher orders are added to the fit. Certainly adding both $x^4$ and $x^5$ terms to the fit ($k'=5$) does not alter the results for the coefficients up to $k=3$ compared to what was obtained for $k=3$, $k'=4$.

\subsection{The role of parametric uncertainty and higher-order uncertainty in predictions}

In this section we derive the 
pdf of our prediction for $g$ at some new point $x$. To do that we carry out a specific instance of the general calculation of the prediction pdf that was discussed in Sec.~\ref{sec:theory}. In the calculation of this section we choose $k'=k+1$. We therefore restrict our considerations to marginalization over one higher-order term, but the generalization to the case where there are several higher-order terms in the discrepancy model is straightforward.

We begin by marginalizing over the value of the $k$th-order theory prediction at the new point, ${\yth}_k(x)$, as well as over the coefficient $a_{k+1}$ that defines the model discrepancy:
\begin{align}
    \pr(g(x) \given  D,k,I) = \int \pr(g(x) \given  a_{k+1}, {\yth}_{k}(x),I) \,
    \pr(a_{k+1}, {\yth}_{k}(x) \given  D,k,I) \, da_{k+1} d{\yth}_k(x).
    \label{eq:pdf_of_g(x)} 
\end{align}
As already noted,
we have assumed that the model discrepancy can be well approximated by the first omitted term:
 \begin{equation}
        \delta {\yth}_k(x) = a_{k+1} x^{k+1},
    \end{equation}
so if
\begin{equation}
    a_{k+1} \sim {\cal N}(0,\abar^2),
    \label{eq:akplus1}
\end{equation} 
 we recover Eq.~(\ref{eq:deltaythnat}). 
 
 In order to work out the ${\yth}_k(x)$ dependence of the integral we employ
 the polynomial form for ${\yth}_k(x;\avec)$ from Eq.~\eqref{eq:EFTOk} and marginalize in
the coefficients $a_0$, $a_1$, \ldots, $a_k$. This yields
\begin{align}
        \pr(g(x) \given  D,k,I) =  \int \pr\Bigl(g(x)\given  \sum_{i=0}^{k+1} a_i x^i, I\Bigr) \,
     \pr(a_{k+1},a_0,a_1,\ldots,a_k\given  D,I) \, da_{k+1} da_0 \ldots da_k, 
  \label{eq:ppdpre}
\end{align}
which can be thought of as the instantiation of Eq.~\eqref{eq:ppdfull} in this particular toy problem. 

Because of Eq.~\eqref{eq:gfk} (and more generally Eq.~\eqref{eq:theory_truth}), the first posterior on the right sides of Eqs.~\eqref{eq:pdf_of_g(x)} and \eqref{eq:ppdpre} is a delta function, i.e.,
\begin{eqnarray}
    \pr(g(x) \given  D,k,I) = \int \delta\Bigl(g(x) - \sum_{i=0}^{k+1} a_i x^i\Bigr) \, 
     \pr(a_{k+1},a_0,a_1,\ldots,a_k\given  D,I) \, da_{k+1} da_0 \ldots da_k.
  \label{eq:ppd}
\end{eqnarray}

Ultimately, we are dealing with a linear model here, so the integrals over $a_0, a_1, a_2, \dots, a_k$ will produce a Gaussian distribution for the random variable $\sum_{i=0}^k a_i x^i$, i.e., for the value of ${\yth}_k(x;\avec)$, which we denote ${\yth}_k$ hereafter.
We use the product rule to rewrite:
\begin{equation}
    \pr(a_{k+1},a_0,a_1,\ldots,a_k\given  D,I)=
    \pr(a_{k+1}|\avec,D,I) \pr(\avec|D,I),
\end{equation}
and then perform the integral over the parameters $\avec$ and replace them by a single integral over a Gaussian distribution for ${\yth}_k$.
\begin{eqnarray}
    \pr(g(x) \given  D,k,I) \propto \int \delta\Bigl(g(x) - {\yth}_k - a_{k+1} x^{k+1}\Bigr) \exp\left(-\frac{({\yth}_k - \langle {\yth}_k \rangle)^2}{2 \sigmaparam(x)^2}\right) \, 
     \pr(a_{k+1}\given \avec,D,I) \, da_{k+1} d{\yth}_k.
  \label{eq:ppd}
\end{eqnarray}
 The width of that Gaussian is what we call the parametric error, $\sigmaparam(x)$, of the prediction. This achieves our intermediate goal of working out the ${\yth}_k$ dependence of the integral \eqref{eq:pdf_of_g(x)}.

Now, in order to trace the consequences of assuming the model uncertainty for our predicted observable is independent of its parametric uncertainty, we take
$a_{k+1}$ to be conditionally independent of $\avec$, i.e.:
\begin{equation}
\pr(a_{k+1}\given \avec, D,I) = \pr(a_{k+1}\given D,I).
\label{eq:indparamposterior}
\end{equation}
We also assume that this one-dimensional pdf for $a_{k+1}$ is equal to the prior---as we saw above, this was approximately true for the data set we are considering:
\begin{equation}
\pr(a_{k+1}\given D,I)=\pr(a_{k+1}\given I).
\end{equation}
For the case of a Gaussian prior we then have:
\begin{eqnarray}
\pr(g(x) \given  D,k,I) \propto \int \delta\Bigl(g(x) - {\yth}_k - a_{k+1} x^{k+1}\Bigr) 
\exp\left(-\frac{1}{2} \frac{({\yth}_k - \langle {\yth}_k \rangle)^2}{2 \sigmaparam(x)^2}\right) \exp\left(-\frac{a_{k+1}^2}{2 \abar^2}\right) \,
  d{\yth}_k \, da_{k+1}.
  \label{eq:ppduncorrelated}
\end{eqnarray}
Comparing with Eq.~\eqref{eq:pdf_of_g(x)} we see what the different pieces there have evaluated to in our calculation, and, in particular that the joint probability of $a_{k+1}$ and ${\yth}_k(x)$ has factorized because we assumed that $a_{k+1}$ was independent of the other $a_i$'s. 

To finish the calculation we define
\begin{equation}
    \sigmaHO(x) \equiv \abar x^{k+1},
    \label{eq:houncertprior}
\end{equation}
and find, on performing the Gaussian integrals of Eq.~(\ref{eq:ppduncorrelated}), that the width of the ppd for $g(x)$ is 
\begin{equation}
\sigma_{\text{full}}(x) = \sqrt{\sigmaHO(x)^{2} + \sigmaparam(x)^{2}}.
\label{eq:sigmafull}
\end{equation}
We have therefore recovered the prescription that higher-order and parametric uncertainties should be added in quadrature. 

But this result ignores the correlations between $a_{k+1}$ and $a_0$, \ldots, $a_k$ that result from fitting the parameters being estimated to data using a discrepancy term in which $a_{k+1}$ appears. In the next section we will explore what happens when we retain those correlations, i.e., evaluate the integral (\ref{eq:ppd}) without assuming Eq.~(\ref{eq:indparamposterior}), i.e., that the $a_{k+1}$ is conditionally independent of $\avec$.

\section{The interplay of uncertainties in predictions}
\label{sec:predictions}

We now compute the pdf of Eq.~(\ref{eq:pdf_of_g(x)}) for the case $k=3$. This, of course, can also be understood as a fourth-order fit to the data. But thinking of the fourth-order effect as truncation uncertainty allows us to see what happens when we switch it on and off, and also when we approximate the way it appears in the PPD.

\begin{figure}[htb!]
\centering
\includegraphics[width=0.8\textwidth]{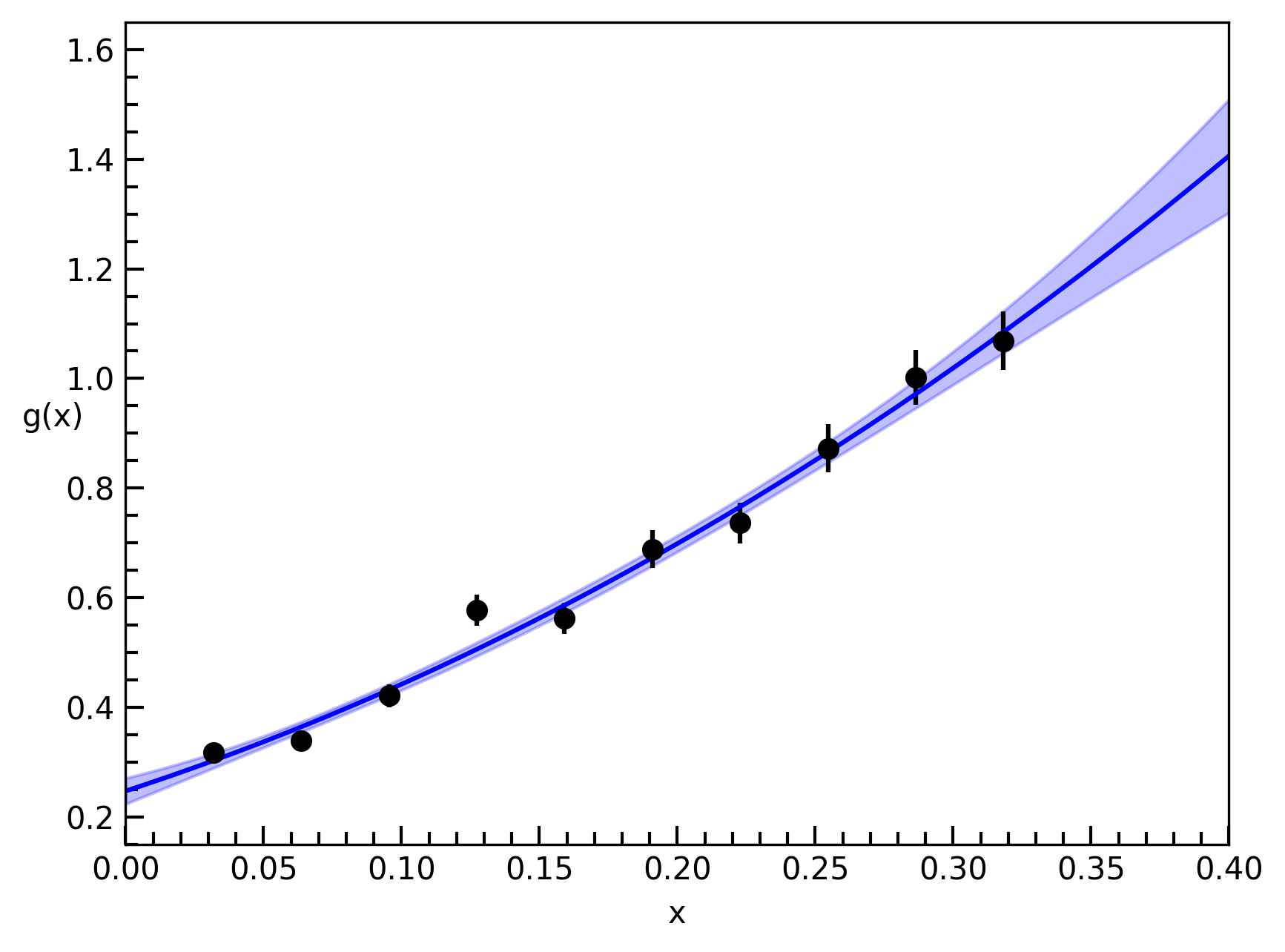}
\captionof{figure}{The prediction for the value of the function $g(x)$ based on Bayesian linear regression carried out with a third-order polynomial, with a fourth-order term treated as model uncertainty and marginalized over. The prediction is conditioned on the data set $D_1$ from Refs.~\cite{Schindler:2008fh,Wesolowski:2015fqa}. The blue band represents the total model uncertainty of the prediction.}
\label{fig:model_uncertainty}
\end{figure}

The result of the full inclusion of the fourth-order effect is shown in Fig.~\ref{fig:model_uncertainty}. 
This gives a narrow ppd for the function in the region covered by the dataset. For values of $x$ above the region covered by data the uncertainty in the prediction gets larger.

\begin{figure}[tbh!]
\includegraphics[scale=0.8]{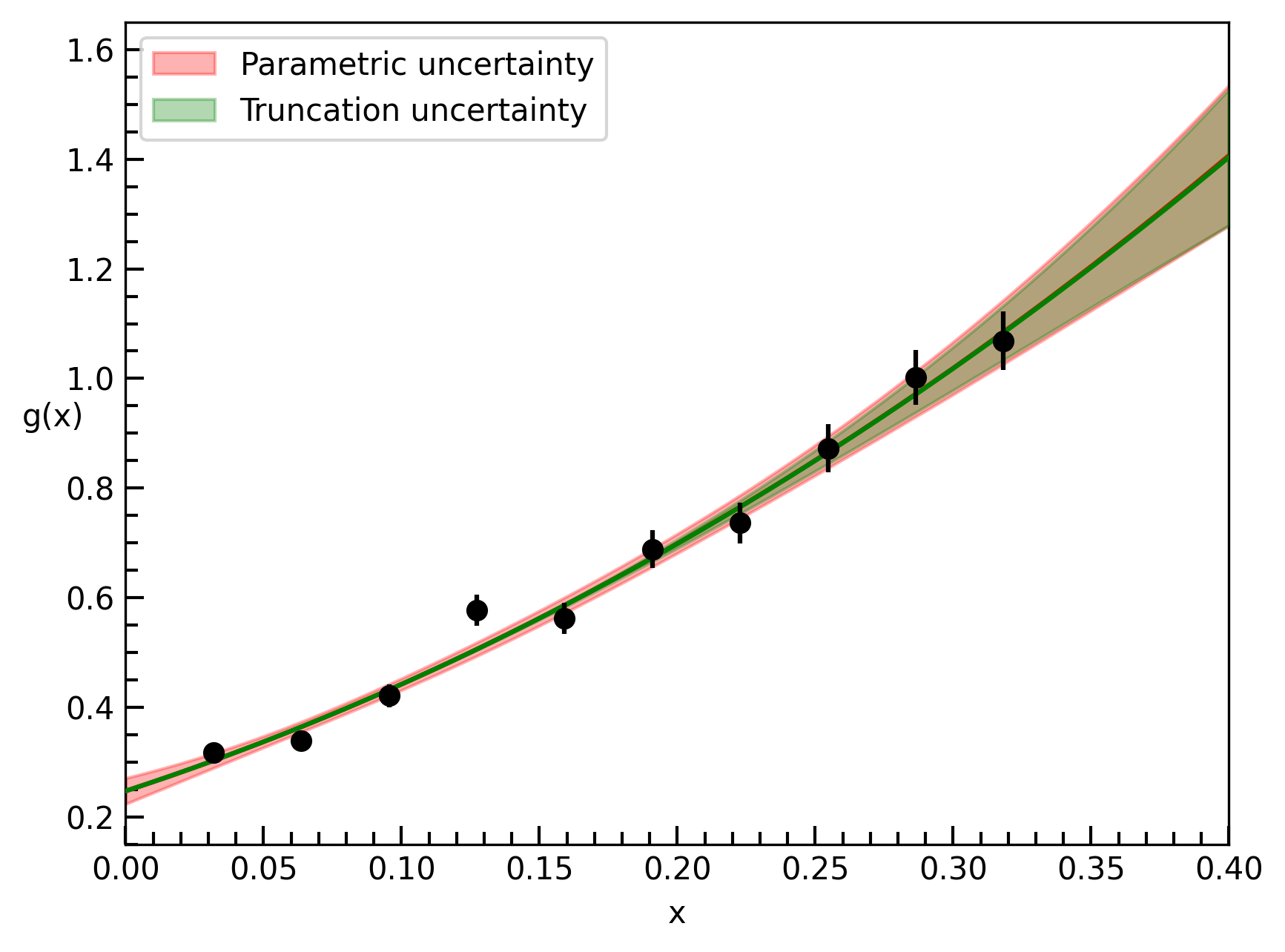}
\centering
\captionof{figure}{Third-order polynomial prediction for the function $g(x)$, conditioned on the data set $D_1$---the same data set as in Fig.~\ref{fig:model_uncertainty}. However, here the parameter uncertainty, $\sigmaparam(x)$, which is depicted as the red band, and the truncation uncertainty, $\sigmaHO(x)$, depicted as the green band, are assessed separately. Comparing to Fig.~\ref{fig:model_uncertainty}, we see that the actual uncertainty associated with the fourth-order term in the polynomial is smaller than either the parametric or truncation uncertainty bands shown here.}
\label{fig:paramvstrunc}
\end{figure}

%
We first compute the parametric uncertainty by setting $a_{k+1}$ to zero. Let us suppose we have $M$ MCMC samples of our parameters $a_0$, $a_1$, \ldots, $a_k$ and the discrepancy term $a_{k+1} x^{k+1}$. We construct the uncertainty at order $k$ by computing the sum of the series to order $k$:
\begin{equation}
 {\yth}_k(x;\avec) \equiv    \sum_{i=0}^k a_i x^i,
\end{equation}
for each of the $M$ samples of the vector $\avec$. We then compute the mean and standard deviation of ${\yth}_k(x;\avec)$ over these $M$ samples. This establishes the central value of ${\yth}_k(x;\avec)$ ($\langle {\yth}_k \rangle$ in Eqs.~\eqref{eq:ppd} and \eqref{eq:ppduncorrelated}) and its parametric uncertainty, $\sigma_{\rm param}(x)$. They are depicted as the red line and the red shaded band in Fig.~\ref{fig:paramvstrunc}. We note that this parametric uncertainty indirectly includes the effect of higher-order coefficients (specifically $a_{k+1}$) on $a_0, \ldots, a_k$, since $a_{k+1}$ was accounted for when the samples of $\{a_0,a_1,\ldots,a_k\}$ were generated.

We also compute the direct effect of higher-order uncertainty, $\sigmaHO$, from the samples as:
\begin{equation}
    \sigmaHO=\sqrt{\bigl\langle a_{k+1}^2 x^{2k+2}\bigr\rangle},
    \label{eq:trunc_error}
\end{equation}
where the average here is taken over the $M$ samples. This is represented by the green band in Fig.~\ref{fig:paramvstrunc}, with the median value of the green band defined by the median value of ${\yth}_k(x;\avec)$. In fact, this green band shows the higher-order effect obtained from ${\rm pr}(a_{k+1}\given D,I)$. That result for Eq.~\eqref{eq:trunc_error} will differ from the formula (\ref{eq:houncertprior}) to the extent that the pdf of $a_{k+1}$ is influenced by the data. If the one-dimensional pdf for $a_{k+1}$ coincides with the prior ${\cal N}(0,\abar^2)$ then using samples of $a_{k+1}$ to construct Eq.~\eqref{eq:trunc_error} will recover Eq.~(\ref{eq:houncertprior}). In the case considered here the one-dimensional pdf of $a_4$ is very close to the prior, see Table~\ref{table:parameters}.

\begin{figure}[tbh!]
\centering
\includegraphics[scale=0.8]{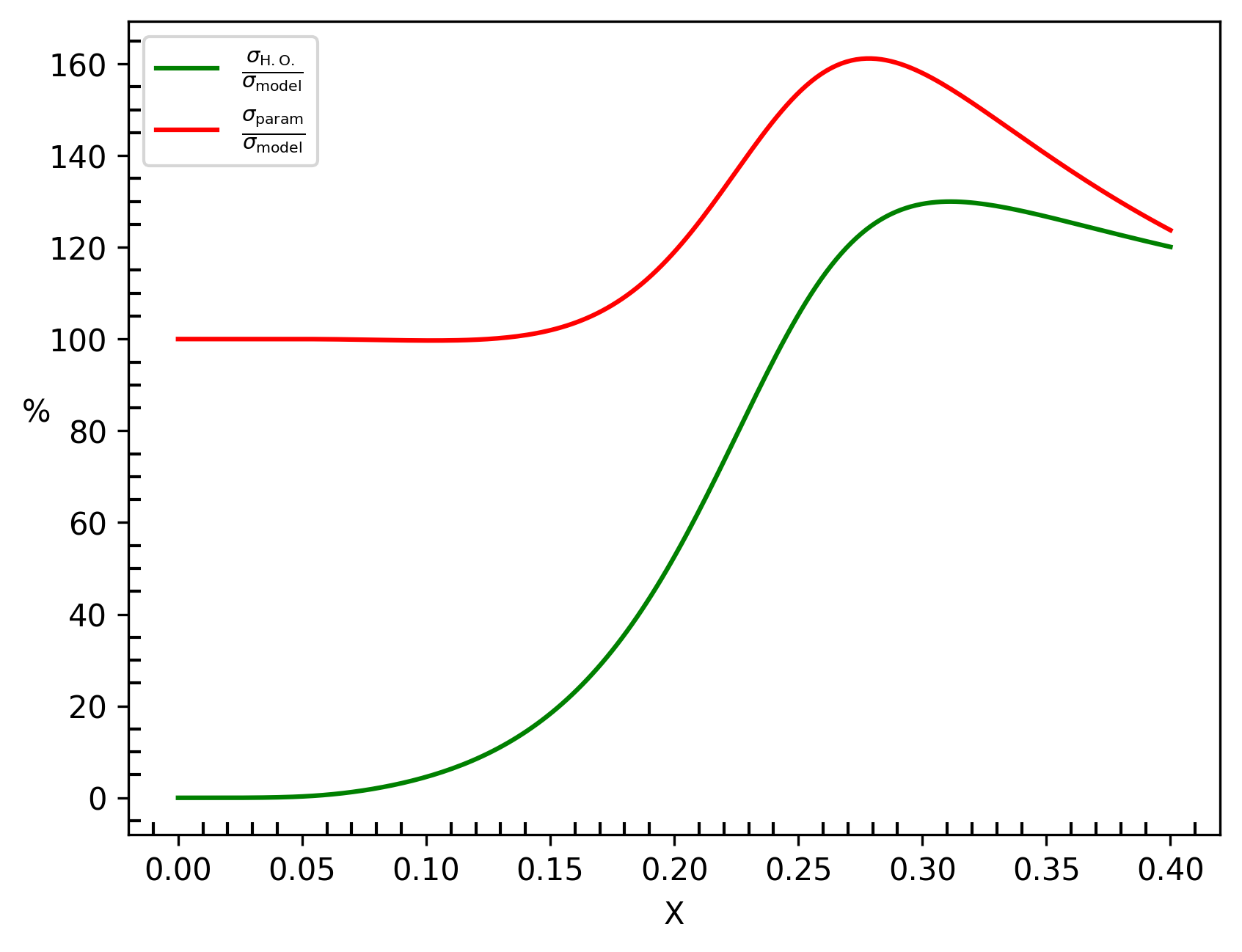}
\captionof{figure}{The red line depicts the ratio of the parametric uncertainty $\sigma_{\rm param}(x)$ to the full uncertainty as a percentage. The green line depicts the ratio of the truncation uncertainty $a_{k+1} x^{k+1}$ to the full uncertainty. }
\label{fig:compare}
\end{figure}

Figure~\ref{fig:paramvstrunc} reveals that the parametric and truncation errors are individually large---indeed, they're a little bit larger than the errors in the data set $D_1$ (see Table~\ref{tab:toy_data}) used to calibrate the model. 

This information is redisplayed as the fractional difference of each uncertainty to the full uncertainty in Fig.~\ref{fig:compare}. At low $x$ the full uncertainty is completely given by the parametric uncertainty. However, for $x \approx 0.25$ the parametric (truncation) uncertainty is 60\% (20\%) larger than the full uncertainty plotted in Fig.~\ref{fig:model_uncertainty}.
This suggests an anti-correlated relationship between the random variables that represent the two different sources of uncertainty.

\begin{figure}[tbh!]
\centering
\includegraphics[scale=0.8]{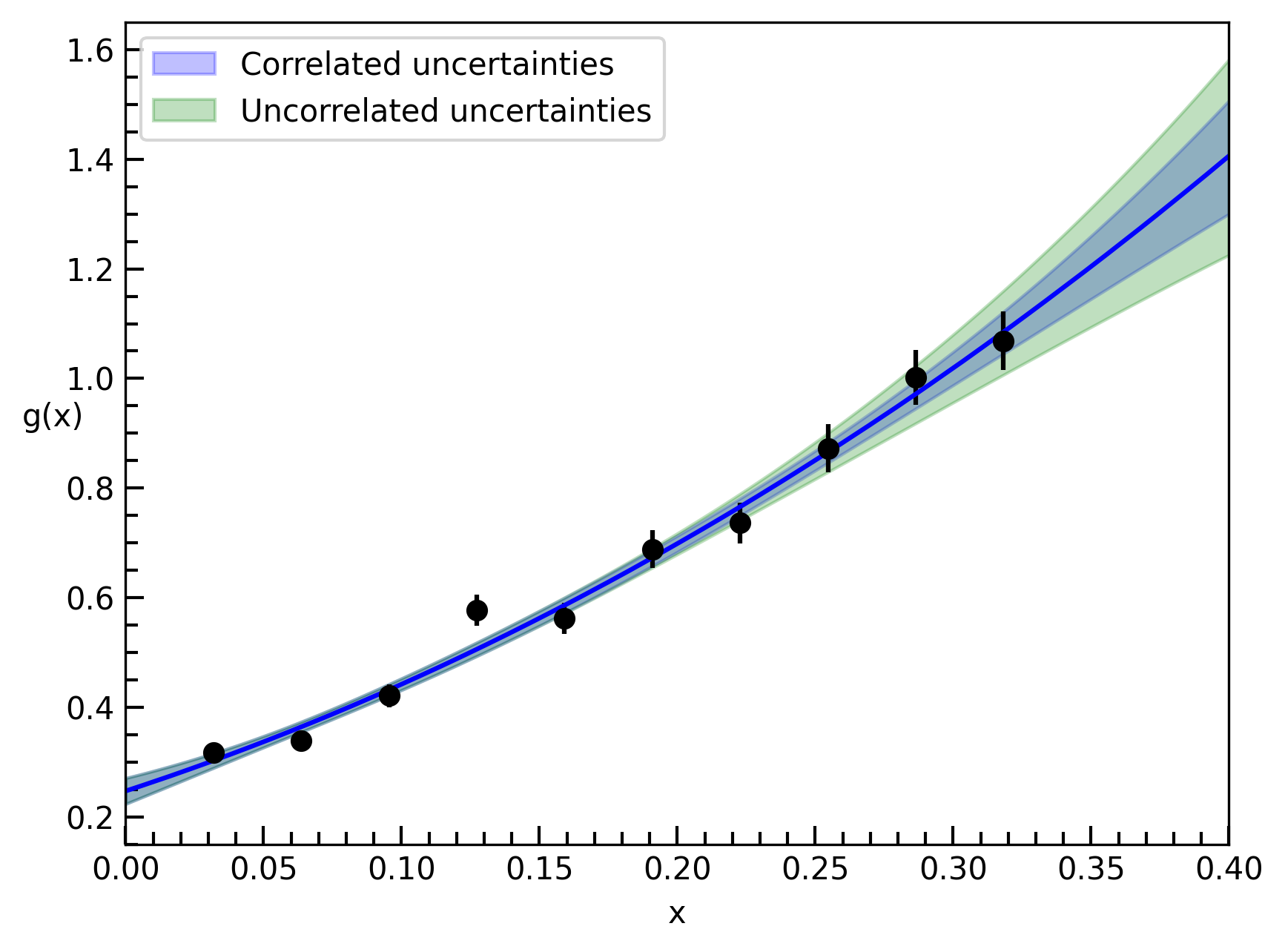}
\captionof{figure}{The full uncertainty of the model, computed by considering the anti-correlation as in Eq.~(\ref{eq:sigmacorrect}), is shown in blue. The result obtained when this correlation is neglected, and so  the truncation error and the parametric error are combined in quadrature, per Eq.~\eqref{eq:sigmafull}, is the markedly wider green band.
The polynomial fit here is up to order $k=3$, and the quartic term is treated as a truncation error.}
\label{fig:errorcombination}
\end{figure}

As discussed in the previous section, if we neglect the correlation between the discrepancy term $a_{k+1} x^{k+1}$ and the estimated parameters $a_0, \ldots, a_k$, then the width of the posterior predictive distribution for $g(x)$ is given by the sum in quadrature of the parametric and higher-order errors, see Eq.~(\ref{eq:sigmafull}). 
This result is represented by the green band in Fig.~\ref{fig:errorcombination}. This uncertainty leads to Bayesian credibility intervals that are too large: the prediction is much closer to the data than such large uncertainties might lead one to expect. The blue band in Fig.~\ref{fig:errorcombination} repeats the full result from Fig.~\ref{fig:model_uncertainty}, and accurately quantifies the uncertainty in the predictions. 

These overestimated uncertainties occur because the truncation uncertainty in our prediction for $g(x)$, as defined by Eq.~(\ref{eq:trunc_error}), is highly anti-correlated with $\yth(x)$'s parametric uncertainty, at least for the domain of input values studied here. The anti-correlation of these two uncertainties is inherited from their anti-correlation in the region of the input space $x$ where the data constrains the model parameters $\avec$. In other words, the likelihood {\it does} constrain $a_{k+1}$: it doesn't constrain its one-dimensional posterior very much (see the second-last column of Table~\ref{table:parameters}), but it forces anti-correlation between $a_{k+1}$ and the LECs $a_0, \ldots, a_k$. This anti-correlation ensures that the variability in $g(x)$, as constructed in Eq.~\eqref{eq:gfk}, is of order $0.05 g(x)$, as specified by Eq.~\eqref{eq:datageneration}. 

Note that in order for the anti-correlation between $a_{k+1}$ and $\avec$ to be inherited by the prediction it matters that there is a large correlation coefficient between the polynomial model $M$'s predictions for $g$ at the new point $x$ and its results at the $x_i$ that were used to calibrate the coefficients.
The limiting factor on the accuracy of $M$'s prediction for $g(x)$ in this region is then neither the parametric nor the model uncertainty, but the statistical power of the data set $D$.

\begin{figure}[tbh!]
\centering
\includegraphics[width=0.8\textwidth]{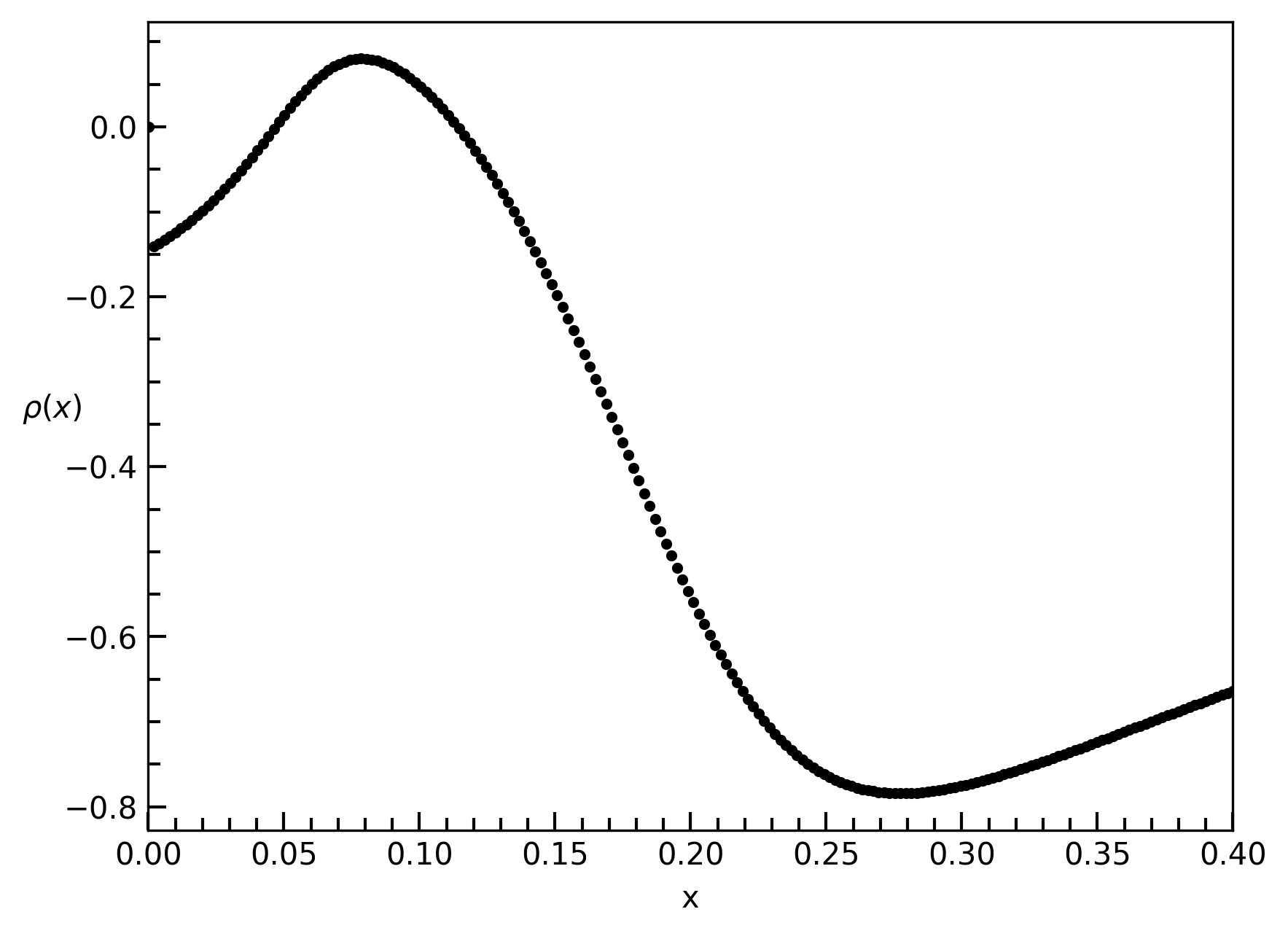}
\captionof{figure}{The Pearson correlation coefficient $\rho(x)$ from Eq.~\eqref{eq:pearson} of the parametric and higher-order uncertainty as a function of $x$.}
\label{fig:pearson}
\end{figure}

To reveal the extent of the anti-correlation between deviations from the mean value of the prediction due to parameter uncertainty and deviations due to higher-order uncertainty we 
 compute the Pearson correlation coefficient of the two sources of fluctuation:
\begin{equation}
\rho(x) =\frac{\Bigl\langle\bigl(f_{k}(x)-\bigl\langle f_{k}(x)\bigr\rangle\bigr) \bigl({\delta \yth}_k(x)-\bigl\langle {\delta \yth}_k(x)\bigr\rangle\bigr) \Bigr\rangle}{\sigmaHO \sigmaparam} = \frac{\Bigl\langle\bigl(f_{k}(x)-\bigl\langle f_{k}(x)\bigr\rangle\bigr) \bigl(a_{k+1} x^{k+1}\bigr) \Bigr\rangle}{\sigmaHO \sigmaparam},
  \label{eq:pearson}
\end{equation}
with ${\delta \yth}_k=a_{k+1} x^{k+1}$.
Here the average is again taken over  the $M$ MCMC samples. The result for $\rho(x)$ is shown in Fig.~\ref{fig:pearson}.

The higher-order and parametric errors are then combined according to
\begin{equation}
\sigma_{\text{model}} = \sqrt{\sigmaHO^{2} + 2\rho(x) \sigmaHO \sigmaparam + \sigmaparam^{2}}.
\label{eq:sigmacorrect}
\end{equation}
Note the difference from Eq.~\eqref{eq:sigmafull}, which was obtained from a factorized pdf in which the (anti-)correlations shown in Fig.~\ref{fig:pearson} were neglected.

The calculation with the contribution from the correlation coefficient recovers the same uncertainty band as when we retain all the information on ${\rm pr}(a_0,a_1,a_2,a_3,a_4\given D,I)$, see Fig.~\ref{fig:model_uncertainty}. In contrast, summing the uncertainties in quadrature (i.e., without the correlation coefficient) results in an overestimation of the prediction's uncertainty. The overestimation may appear modest in Fig.~\ref{fig:errorcombination}, but the uncertainty is as much as a factor of two too large even in this simple example. And for the case of a bigger $\abar$, or smaller data uncertainties, the failure to account for the anti-correlation of the two sources of higher-order uncertainty that is induced by the calibration of parameters to data in the presence of discrepancy becomes quite dramatic. If, in the above example, we reduce the data uncertainties to 1\%, then assuming independence of parametric and truncation uncertainties leads to uncertainties in predictions that are a factor of eight too large.

\section{Conclusions}
\label{sec:conclusions}

Bayesian methodology permits quantitative assessment of the impact that model discrepancy---often called ``model uncertainty''---has on parameter estimation. Model discrepancy should also be considered when predicting as-yet-unmeasured quantities with the model. However, neither model discrepancy nor parameter uncertainty alone can give an accurate picture of the full uncertainty of a model prediction. Somewhat counterintuitively, considering only one or the other is likely to lead to {\it over}estimation of the model uncertainty, at least if the quantity being predicted has any significant correlation with the data that were used to calibrate the parameters of the model. 

In the context of effective field theories (EFTs) the model discrepancy is constrained in regard to both its parametric dependence on the input space and its size. In this paper we examined a toy model of EFTs in which Bayesian linear regression for a polynomial of order $k$ is performed on data, and the impact of terms $\sim x^{k+1}$ and higher is accounted for in the Bayesian posterior. 

If all that is desired is a posterior for the coefficients $a_0, \ldots a_k$, then the coefficients $a_{k+1}, \ldots$ can be marginalized out of the calculation. In previous work we have shown the effect of these higher-order parameters can also be accounted for by modifying the likelihood. The covariance matrix that appears there should be the sum of the usual experimental covariance and a theory covariance matrix, $\Sigma_{\rm th}$, that encodes the effect of truncation error on parameter estimation~\cite{Wesolowski:2018lzj}. 

Here we showed that making predictions with the LEC posterior ${\rm pr}({\bf a}\given D,I)$ that results from such a modified likelihood and separately including the truncation error will lead to overly wide credibility intervals unless the correlation structure induced by the calibration procedure is taken into account. The deviation from data that a particular set of parameters produces is anti-correlated with the truncation error. If we retain the information on the model discrepancy/truncation error in the posterior pdf, that is, compute ${\rm pr}(a_0,\ldots,a_k,\delta y_{\rm th}\given D,I)$, and use this joint posterior of the model parameters and the discrepancy term for prediction at a new point, the truncation uncertainty will be properly accounted for in those predictions. 

In Ref.~\cite{Brynjarsdottir:2014} Brynjasdottir and O'Hagan examined the impact that model discrepancy has on both model parameters and predictions. In contrast to our EFT-based study, they used a Gaussian process to describe $\delta y_{\rm th}$. Nevertheless, they found that there was a strong anti-correlation between the parameter random variable and the model discrepancy. Failure to consider model discrepancy produced overly confident parameter inference and poor predictions at large input values. Accurate predictions require the retention of information on the correlations between the model discrepancy and the parameter(s). Uncertainty can be properly calculated if information on these correlations is retained in the Bayesian framework.

\acknowledgments{We thank Andreas Ekstr\"om, Harald Grie\ss hammer, Sunil Jaiswal, and Adam Murphy for useful conversations. This work was supported by the US Department of Energy under contract no. DE-FG02-93ER-40756 (DRP) and DE-SC0015393 (NC) and the NUCLEI SciDAC Collaboration under contract DE-FG02-96ER40963 (RJF), and by the National Science of Foundation via CSSI program Award no. OAC-2004601 (BAND collaboration \cite{BAND_Framework}, DRP \& RJF) and via Award Nos.~PHY-2209442 and PHY-2514765 (RJF).}

\bibliography{bayesian_refs}
\end{document}